\def\gsim{\;\lower 0.5ex\hbox{$\buildrel > \over \sim\ $}}
\def\lsim{\;\lower 0.5ex\hbox{$\buildrel < \over \sim\ $}}
\documentstyle{lamuphys}
\makeatletter
\let\chapter\hid@chapter
\makeatother
\begin{document}
\pagenumbering{arabic}
\title{Molecular Gas in Quasar Hosts$^1$}

\author{Richard\,Barvainis}

\institute{MIT Haystack Observatory, Westford MA USA 01886}

\maketitle

\addtocounter{footnote}{1}
\footnotetext{Invited Review, to appear in {\it Quasar Hosts}, proceedings
of the ESA-IAC conference held on Tenerife, Canary Islands, 24--27 Sept 1996, 
ed. D. Clements and I. Perez-Fournon [Springer-Verlag].}

\begin{abstract}
The study of molecular gas in quasar host galaxies addresses a number
of interesting questions pertaining to the hosts' ISM, to unified schemes
relating quasars and IR galaxies, and to the processes fueling nuclear 
activity.  In this contribution I review observations 
of molecular gas in quasar hosts from $z = 0.06$ to $z = 4.7$.  
The Cloverleaf quasar  at $z = 2.5$ is featured as a case where there are now 
enough detected transitions (four in CO, and one each in CI and HCN) to allow 
detailed modeling of physical conditions in the molecular ISM.
We find that the CO-emitting gas is warmer, denser, and less optically
thick than that found in typical Galactic molecular clouds.  These 
differences are probably due to the presence of the luminous quasar in the 
nucleus of the Cloverleaf's host galaxy.

\end{abstract}
\section{Introduction}

  Prior to the flight of the IRAS satellite in the mid-1980s, it was
assumed by most workers in the field that quasar infrared emission
was nonthermal in character, for both radio quiet and radio loud 
quasars.  The broad-band view of the IR provided by IRAS showed
that the continua were not necessarily well-described by a simple 
power law, although in some instances such a description could suffice.
The question of whether the IR might rather be dust emission was 
then brought to the fore, with many implications for the IR itself, 
and also for the nature of the energetically
dominant optical/UV Big Blue Bump.  One potential 
discriminant between the nonthermal (synchrotron) and dust models
would be the presence of CO emission commensurate with the far-IR 
strength, and this test was what first led some of us to look for CO from 
quasars back in the late `80s.

  CO was in fact detected in amounts expected from the dust model, and 
this provided one avenue of evidence among many leading to the current widely
accepted view that emission from dust is the dominant IR process in most 
radio quiet and normal (i.e., non-blazar) radio loud objects (see Barvainis
1992 for a review).   This having been established, there are a number
of other interesting and important questions that can be addressed by
the study of molecular gas in quasars.  They include:
\begin{itemize}  

\item How is the host ISM affected by the presence of the AGN?
\item Do quasar hosts have enhanced ISM as a result of interactions 
and mergers?
\item How are quasars and luminous IR galaxies related, and how 
are they fueled?
\item What are the properties of star formation at high redshift?
\item And, speculatively, can we use CO to measure redshifts of obscured 
quasars at ultra-high $z$?

\end{itemize}

  Here I review observations to date of molecular gas in quasar hosts and 
discuss some of what has been learned, with a focus on the Cloverleaf quasar
where we now have enough transitions to perform detailed modeling 
of the molecular gas within the inner few hundred parsecs of the active galactic
nucleus.

\section{Overview of Observations}

\subsection{Low Redshifts}

  The molecular species with the strongest lines at radio wavelengths is CO, 
a diatomic molecule with rotational transitions spaced by 115 GHz.  
Initial searches 
focussed on the $J=1\rightarrow 0$ transition in low redshift quasars 
detected at 60 and/or 100~$\mu$m by IRAS.  CO had already been
seen in abundance in luminous IR galaxies (now believed
by many to harbor hidden quasar nuclei that have been both activated and
obscured by processes accompanying galactic interactions and mergers 
- see Sanders and Mirabel 1996 for a review).  

Sanders, Scoville, \& Soifer (1988)
reported detection of CO(1--0) emission from the infrared-excess quasar
Mrk 1014 at $z = 0.16$, with a total molecular gas mass $M_{\rm H_2}
\sim 4\times 10^{10} M_{\sun}$ (using a conversion 
factor from CO luminosity to H$_2$ mass appropriate for Galactic molecular
clouds; however, this value may be 
too large for IR galaxies and quasars -- see below).  
Barvainis, Alloin, \& Antonucci (1989) detected strong lines in both
CO(1--0) and CO(2--1) from the $z = 0.06$ quasar I Zw 1, which showed 
double-horned profiles very similar to the HI profile from this
system.  Sanders et al (1989) detected CO(1--0) from the IR-discovered
quasar IRAS 07598+6508.  CO detections from several other low-$z$ quasars 
were reported by Alloin et al (1992), lending credence to the dust model
for quasar IR emission.  In contrast to the objects mentioned so far, 
which are all radio quiet, Scoville et al (1993) detected CO from 3C48,
a radio loud quasar at $z = 0.3$ with a massive elliptical host showing 
signs of a recent merger (Stockton and Ridgway 1991; Barvainis 1993).

\subsection{High Redshifts}

  The first claimed detection of molecular gas at high redshifts was the 
CO(3--2) line from IRAS F10214+4724, an ultraluminous IR galaxy at 
$z = 2.28$, by Brown and Vanden Bout (1991).  Although
the original report turned out to be incorrect (Radford et al 1996), a 
follow-up study by Brown and Vanden Bout (1992) did correctly measure
a (weaker) line in CO(3--2), and also a line in CO(4--3);  the CO(3--2)
line was confirmed at many telescopes world-wide (see Radford et al for 
references).
CO(6--5) has been detected as well, by Solomon, Radford, \& Downes (1992),
but a claimed detection of CO(1--0) by Tsuboi \& Nakai (1994) was not 
confirmed by Barvainis (1995).

  Today we know two important facts about F10214+4724 which were not known
in 1992:  It is gravitationally lensed (e.g., Broadhurst and Lehar 1995), 
and it contains a hidden quasar in its nucleus (Goodrich et al 1996).
This makes it similar in several ways to the second high redshift object 
to be discovered in CO, the Cloverleaf quasar at $z = 2.56$
(Barvainis et al 1994), to which we turn in the next section for a more 
detailed look.  After accounting for lensing magnification, current 
estimates for the molecular gas mass in F10214+4724
have been scaled down considerably relative to earlier reports, but
are still fairly large: $M_{\rm H_2} \sim 2\times 10^{10}~ M_{\sun}$  
[using the Galactic value of 
$M_{\rm H_2}/L'_{\rm CO} = 4.0 M_{\sun}$ (K km s$^{-1}$ pc$^2$ )$^{-1}$].
This estimate is based on interferometric observations by Downes, Solomon, \& 
Radford (1995), who also find that the bulk of the CO is confined to the 
inner part of the galactic nucleus ($r_{\rm CO} \leq 400-800$ pc).  
These authors conclude that this object is not very different from luminous 
IR galaxies in the local Universe.

  At the very highest redshifts known, a very recent and remarkable 
discovery has been made of CO(5--4) emission from the quasar BR1202--0725, at
$z = 4.7$.  Simultaneous papers in Nature by Ohta et al (1996) and 
Omont et al (1996) confirm the reality of the CO(5--4) detection, while 
Omont et al also show a probable detection of CO(7--6).  
The Omont et al observations, with $2''$ resolution, were able to resolve
the CO source into two spatially distinct components, separated by
$4''$.  One component is precisely coincident with the optical quasar,
but the second component has no optical counterpart on deep HST images.
This second component is either a gravitationally lensed image in which
the optical part is obscured by passage through a dust cloud in the 
(as-yet-undetected) lensing galaxy, or it could be a real, optically-weak,
molecular companion.  The latter interpretation is by far the more 
interesting one, for reasons I will touch upon in the final section of 
this review.

\subsection{No Redshifts}

  All of the quasars detected in CO so far have been selected for having
either far-IR detections from IRAS, or submillimeter detections from 
ground-based
observations in the 400--1300~$\mu$m region. The submm dust spectrum is 
rising so steeply toward shorter wavelengths that at high redshifts 
($z \gsim 1$) the favorable K-correction 
keeps up with geometrical dilution, so that more distant sources are no 
harder to detect in the submm continuum than nearer ones.  Detectable 
far-IR/submm 
flux (using current instruments) seems to be a necessary but not sufficient 
condition for CO detections, and is the only known selection criterion 
for CO that does in fact work at least part of the time.  
%detection limits for far-IR continuum and CO line emission from a given 
%object are comparable.

  Searches based on other criteria have not been successful.  For example,
Barvainis \& Antonucci (1995) selected most of the known $z \approx 4$ quasars
for which the 115 GHz CO(1--0) line frequency is shifted into the 
VLA 22--24 GHz band.  They failed to obtain detections at typical levels of 
$L'_{\rm CO} \lsim 10^{11}$ K km s$^{-1}$ pc$^2$ for 10 objects.  
Evans et al (1996) derived limits up to an order of magnitude lower in some 
cases for a sample of 11 powerful radio galaxies lying in the redshift range 
$1 < z < 4$,  only one of which had a thermal far-IR/submm detection. 
Brown \& Vanden Bout (1993) reported detection of CO(3--2) emission at 
$z = 2.14$
from a damped Lyman--$\alpha$ absorption system towards the quasar Q0528--250, 
and Frayer et al (1994) reported CO(1--0) and CO(3--2) 
from a DLA at $z = 3.137$.
Both of the CO(3--2) DLA results were refuted by more sensitive observations,
in which a number of other DLA systems were also searched for CO 
and not detected (Wiklind \& Combes 1994; Braine, Downes, \& Guilloteau 1996).

  We should not leave this section on nondetections without mentioning
that the redshift to be used for tuning one's receiver for molecular line
observations of quasar hosts should be based 
on LOW IONIZATION broad optical emission lines,
e.g.\ H$\alpha$, Mg II, etc., or narrow forbidden lines like [OIII].   
`High ionization' lines such as Lyman--$\alpha$ or CIV can be blueshifted
with respect to systemic by as much as several thousand km s$^{-1}$, and
tunings based on such lines will often result in incorrect placement
of narrow mm-wavelength spectrometers. 

\section{The Nuclear ISM in the Cloverleaf Quasar}

  The Cloverleaf is a broad absorption line quasar at $z = 2.56$, 
gravitationally lensed into four spots in the optical with separations  
of about $1''$.  The total magnification factor is estimated
to be $\sim 10$.  

  Our initial search for molecular gas in the Cloverleaf was motivated
by a detection of strong submm flux during a survey of BALQs
using the JCMT (Barvainis, Antonucci, \& Coleman 1992).
After two missed attempts at CO(3--2) using the IRAM 30m telescope, caused 
first by a bad redshift (based on Lyman--$\alpha$, which in this object is 
blueshifted by 
$\approx 900$ km s$^{-1}$ relative to systemic), and then
by a mistuned receiver, a strong detection was finally obtained at the 
IRAM Plateau de Bure Interferometer and confirmed using the IRAM 30m telescope
(Barvainis et al 1994).  From follow-up observations at the 30m we now have 
three
additional CO transitions ($J = 4\rightarrow 3$,~~ $5\rightarrow 4$, and
$7\rightarrow 6$), a detection of the neutral carbon fine-structure
transition CI($^3P_1-{^3P_0}$), and a probable detection of HCN(4--3).

  Given the four observed CO lines, it is possible to proceed with 
detailed modeling of the CO-emitting gas based on the line ratios 
and other constraints.  The line brightness temperature ratios relative
to the strongest line [CO(4--3)] are:
$(3-2)/(4-3) = 0.83\pm0.16$, $(5-4)/(4-3) = 0.73\pm0.16$, and 
$(7-6)/(4-3) = 0.68\pm0.13$.  The roughly constant $T_B$ from (3--2) to 
(4--3), followed by a falloff in the higher-$J$ transitions, suggests
that the optical depths in the CO lines are only modest ($\tau_{4-3}
\lsim 3$), and that the gas is relatively warm ($T \gsim 100$ K)
and dense ($n_{H_2} \gsim 3\times 10^3$ cm$^{-3}$).  The gas may be  
heated by X-rays from the quasar, and the optical depths in each 
transition are lowered because of the distribution 
of population over many states at high temperature.  These conclusions
are based on escape probability modeling by Phil Maloney, and are
presented along with the line measurements in Barvainis et al (1997).

  Because of the high temperature and low optical depth of the gas,
it has a much higher emissivity per unit mass than is typical of 
Galactic molecular clouds.  This means that application of 
the standard Galactic $L'_{\rm CO}$ to $M_{\rm H_2}$ conversion 
factor [4 $M_{\sun}$ (K km s$^{-1}$ pc$^2$)$^{-1}$; e.g., Radford, Solomon, 
\& Downes 1991] would overestimate   
the total molecular mass, according to our model, by about an order 
of magnitude.
Our derived mass for the Cloverleaf is $M_{\rm H_2} \approx 
2\times 10^{10}m^{-1}h^{-2} M_{\sun}$, 
where $m$ is the lensing magnification 
factor and $h$ is Hubble's constant in units of 100 km s$^{-1}$ Mpc$^{-1}$.
For $m = 10$ and $h = 0.7$ this 
represents a relatively modest mass of roughly $4\times 10^9 M_{\sun}$, based on
the {\it observed} CO emission (it is possible that there is significant 
mass outside the X-ray heated zone that would escape 
detection because the gas is cool and optically thick).  

  A conflict has been noted in luminous IR galaxies between the molecular 
mass derived from mm-wavelength CO emission line observations,
and the dynamical mass derived from central stellar velocity dispersions
(obtained from near-IR CO absorption bands).
Shier, Rieke, \& Rieke (1994) find that there must be 4--10 times less
molecular mass in NGC 1614 and IC 694 than estimated using the Galactic
conversion factor.   A factor of 10 below the Galactic value is, perhaps 
not coincidentally, the conversion we derived for the Cloverleaf. 

  Is the molecular mass in the Cloverleaf consistent with the
dynamical mass?  We have recently obtained a high resolution ($0.5''$)
image of CO(7--6) and find an upper limit to the diameter of image
C of the quad of 
$< 0.25''$ (Alloin et al 1997, A\&A, in press), yielding an intrinsic 
source radius of $< 2.4 m^{-1} h^{-1}$ kpc ($q_0 = 0.5$).  For $m = 10$ 
and $h = 0.7$, this gives $r < 0.34$ kpc. Combining this 
source size with the  CO line width of 375~km~s$^{-1}$ yields
$$M_{\rm dyn} \approx {r\Delta V^2_{\rm FWHM}\over{G {\rm sin}^2i}}
\lsim  2\times 10^{10} M_{\sun},$$
where the inclination $i$ has been taken to be 45$^{\circ}$.  This is 
comfortably above the derived H$_2$ mass of $4\times 10^9 M_{\sun}$ 
(for the same $m$ and $h$). 

  To summarize this section, the detectability of the Cloverleaf in molecular
and neutral atomic transitions is largely due to high emissivity of the gas
and magnification from gravitational lensing, rather than an extremely large 
mass of gas. 

\section{The Connection Between Quasars and ULIRGs}

The idea that the impressive infrared luminosities ($L_{\rm IR} \gsim 
10^{12} L_{\sun}$) of ultraluminous IR galaxies (ULIRGs) might be 
supplied, or at least augmented, by a hidden quasar goes back to 
shortly after their discovery by IRAS (e.g., Antonucci \& 
Olszewski 1985).   Sanders et al (1988)
expanded this idea to encompass an evolutionary link between ULIRGs
and quasars, with an interaction or merger between two galaxies driving 
a large amount of gas and dust into the galactic nucleus (e.g. 
Barnes \& Hernquist 1996), creating a quasar by providing fuel for a resident
supermassive black hole.  The quasar would be
hidden at first, and the object would appear as a ULIRG.  The IR luminosity
might be supplied by accompanying starbursts and reprocessing of the
AGN UV/X-ray emission via dust.  Later, as
the gas and dust were consumed or dispersed, a classical optical quasar 
would emerge.

  The gas and dust phase, lasting from before the emergence of the quasar until
sometime after, should produce strong molecular line emission.  Indeed,
ULIRGs do of course show copious CO emission (see Sanders \& Mirabel 1996), 
as do quasars with strong far-IR emission  
(see above).  In this context, IRAS F10214+4724 and the Cloverleaf make a 
very interesting pair 
for comparison.  Their redshifts are similar, their IR spectra are almost
identical, and their CO emission properties are very similar.  The two 
objects differ primarily in the optical/UV region, where the Cloverleaf shows
a fairly typical (perhaps slightly reddened) big blue bump, and broad
emission lines, whereas F10214+4724 has a weak and steep optical continuum
and a Seyfert 2 spectrum in unpolarized light.  However, in {\it polarized}
(i.e., scattered) light, F10214 shows broad emission lines (Goodrich et al
1996), confirming that a  
hidden quasar lies in the nucleus (and can probably be seen unobscured from
some other vantage point).  

   In comparing these properties we (Barvainis et al 1995) have suggested
that F10214 and the Cloverleaf may in fact be the same type of object. Both 
would be emerging
from the obscured phase, with a thick and extensive disk or torus of 
material  
still present in the nucleus.  The differences in their observed properties
are likely due to orientation effects:  we see the Cloverleaf from above
the torus and F10214 from the side.   The nuclear light in F10214
is probably scattered by electrons, or perhaps dust clouds, located in 
the opening of the 
torus, and is then reddened by passage through the upper parts of the torus. 
The observed molecular line emission originates in this still large 
reservoir of dust and gas, evident via other 
means in both objects.  In this general picture of ULIRGs and quasars,
molecular gas plays a critical role as the primary fuel for the nuclear
activity, both for starbursts and for the AGN.

\section{Future Prospects}

   High redshift molecular line observations strain current instrumentation
to its limits, but the recent spectacular detection of CO at $z=4.7$ 
shows that continued searches are certainly worth pursuing.  Of the
three known CO-emitting objects at $z > 0.5$, two (F10214+4724 and the
Cloverleaf) are definitely lensed, while the third (BR1202--0725)
has a double CO source, with both CO components having submm counterparts but 
only one having an optical
counterpart.  Determining whether this object is lensed will 
be difficult but critical to evaluating the amount of molecular
mass in high redshift systems, and the future potential for 
observing CO and other molecules at high ($z = 1-5$) 
and possibly ultrahigh ($z > 5$)
redshifts.  Perhaps molecular studies can currently only be done on magnified
systems -- which will limit the number of candidates, but will not 
alter the method of selection by submm continuum, since that component
will be magnified too.   

  If, however, BR1202--0725 is not lensed, it may represent a population
of extremely gas rich systems in the early Universe.  
The companion CO source is particularly intriguing in this regard 
because it has no optical signature, raising the possibility that an 
abundant population of 
dusty molecular objects could have heretofore escaped detection.
Is the sharp falloff in the optical quasar population for $z \gsim 3$
a real decline, or are many early quasars simply obscured by the 
products of massive, galaxy-forming starbursts?  Such objects might 
appear observationally very much 
like the ultraluminous IR galaxies in the local Universe, many of which
are known to host hidden quasars (e.g., Wills et al 1992; Hines et al 1995).   

One argument that has been made
against the obscured quasar hypothesis is based on the near-complete optical 
identification of radio-selected samples (e.g., Shaver et al 1996), but
this conclusion applies mainly to obscuration by intervening absorbers
along the line-of-sight,
and does not appear to strongly constrain internal absorption.
Furthermore, we know that radio-loud objects are found in elliptical
hosts (as was frequently confirmed at this meeting), whereas local
ULIRGs are predominantly radio quiet objects in heavily disturbed systems. 
The proposed hidden quasars at high $z$ might well be similar to the latter,
and such systems
may simply present the wrong type of environment for the production of powerful
radio sources.  Therefore arguments based on identification of radio loud 
samples might not apply.   It should be noted that Webster et al (1995)
have asserted, based on a wide range in B--K colors found in a radio-selected
sample, that up to 80\% of radio loud and radio quiet quasars may be 
reddened enough to be missed
by the usual optical/UV search methods.  However, this result has
been disputed as being due to effects other than reddening by dust (Sergeant \& 
Rawlings 1996; Boyle \& di Matteo 1995; Srianand 1997, this volume; 
Gonzalez--Serrano 1997, this volume), and it appears the jury is still out 
on this point.

  If there {\it are} dusty molecular optically-obscured quasars at ultra-high
redshift, how might we find them?   Random-field surveys for submm sources
are being planned for the new SCUBA camera on the JCMT, and these surveys
may turn up a candidate population based on crude spectral energy distributions
(3--4 points).   But without optical emission lines, it will be 
difficult to obtain accurate redshifts.  An alternative might be 
to determine redshifts via CO observations.   At high redshifts the CO 
rotation ladder is compressed in frequency by the factor $1/(1+z)$, so 
that for say $z = 8$ there would be a CO line every 12 GHz.  Identifying
two such lines would uniquely provide the object's redshift. 
Such an approach, however,  would be impractical using current instruments.  
The next significant advance in CO-detection technology will be the 100m 
Green Bank Telescope (GBT), scheduled for completion in 1998.  In the 35--50
GHz range it will be able to go several times deeper than the IRAM 30m
telescope or the PdBI and OVRO interferometers can at 3mm (and if, at some 
point in the future, the GBT is able to operate at 80--100 GHz as planned, 
it will be 
even more sensitive).  Additionally, the GBT will have a wide instantaneous 
bandwidth of at least 2 GHz, so that stepping through frequency in a search  
for lines will be more efficient than possible now.    

  The combination of SCUBA and the GBT should prove powerful for molecular
studies of known
high redshift quasars as well.  SCUBA detections in the submm are likely to 
provide a good selection of candidates with known redshifts for deep 
integrations by the GBT.  A new 50m telescope for 3mm and 1mm operation,
the Large Millimeter Telescope (LMT), is being built by a US/Mexico
collaboration,
and should be an impressive high-$z$ CO machine when finished in 
$\sim 5$ years.   In the longer term 
($\sim 10$ years) the next step in sensitivity will come with
the new millimeter arrays being planned  
by the US and Japan (especially if combined), and by Europe, which, by 
observing at 2mm or 1mm, will beat the GBT by factors of several.
This progress should be just about enough to match the lensing 
advantage enjoyed by objects like F10214+4724 and the Cloverleaf, 
thereby allowing
molecular studies of high redshift quasars and galaxies having luminosities 
comparable to those of `ordinary' ultraluminous IR objects found in the
local universe.   

%\begin{figure}
%\vspace{2.5cm}
%\caption{The height of this figure is 2.5\,cm}
%\end{figure}

%

%
% ---- Bibliography ----
%

\end{document}